\documentclass[manuscript]{aastex}

\usepackage{natbib}
\usepackage{arydshln}

\shorttitle{Interstellar C$_{60}^{+}$}
\shortauthors{Walker et al.}

\begin{document}

\title{Identification of more interstellar  C$_{60}^{+}$ bands}

\author{ G.\,A.\,H.\,Walker\altaffilmark{1,*}, D.\,A.\,Bohlender\altaffilmark{2}, J.\,P.\,Maier\altaffilmark{3}, E.\,K.\,Campbell\altaffilmark{3}}

\affil{$^1$ 1234 Hewlett Place, Victoria, BC V8S 4P7, Canada; \email{gordonwa@uvic.ca}}

\affil{$^2$ National Research Council of Canada, Herzberg Astronomy and Astrophysics,\\5071 West Saanich Road, Victoria, BC V9E 2E7, Canada; \email{david.bohlender@nrc-cnrc.gc.ca}}

\affil{$^3$ Department of Chemistry, University of Basel, Klingelbergstrasse 80, CH-4056 Basel, Switzerland; \email{j.p.maier@unibas.ch, ewen.campbell@unibas.ch}}

\altaffiltext{*}{Based on observations obtained at the Canada$-$France$-$Hawaii Telescope (CFHT) which is operated by the National Research Council of Canada, the Institut National des Sciences de l'Univers of the Centre National de la Recherche Scientifique of France, and the University of Hawaii.}

\begin{abstract}
Based on gas-phase laboratory spectra at $6\,\textrm{K}$,  \citet{Campbell2015} confirmed that the diffuse interstellar bands (DIBs) at 9632.7 and 9577.5\,\AA\/ are  due to absorption by the fullerene ion C$_{60}^{+}$. They also reported the detection of two other, weaker bands at 9428.5 and 9365.9\,\AA\/. These lie in spectral regions heavily contaminated by telluric water vapour lines. We acquired CFHT ESPaDOnS spectra of HD 183143 close to the zenith and chopped with a nearby standard to correct for the telluric line absorption which enabled us to detect a DIB at 9365.9\,\AA\/ of relative width and strength comparable to the laboratory absorption. There is a DIB of similar strength and FWHM at 9362.5\,\AA\/.  A stellar emission feature at 9429\,\AA\/  prevented detection of the 9428.5\,\AA\/ band. However, a CFHT archival spectrum of HD 169454, where  emission is absent at 9429\,\AA\/, clearly shows the 9428.5\,\AA\/ DIB with the expected strength and width. These results further confirm  C$_{60}^{+}$ as a DIB carrier. 
\end{abstract}


\section{Introduction}
Hundreds of diffuse interstellar bands (DIBs) are seen in the spectra of stars dimmed and reddened by interstellar dust \citep{Her95}. There have been many unsuccessful attempts to assign neutral or charged molecular carriers to either individual or families of these bands. \citet{Foing1994} proposed that a pair of DIBs which they had detected at 9632 and 9577\,\AA\/  were due to the fullerene ion C$_{60}^{+}$ because of their proximity to laboratory absorption wavelengths observed in a neon matrix \citep{Ful93}. \citet{Campbell2015} have now proved that these DIBs are indeed due to  C$_{60}^{+}$. This is based on measurement of not only the wavelengths, but also the FWHM and relative intensities of these bands in the gas phase at $6\,\textrm{K}$. This successful identification, which came long after Kroto first predicted the presence of $\textrm{C}_{60}^{+}$ in the interstellar medium \citep{Kro87}, may be regarded as a first step towards unlocking the 90 year old mystery of the DIB carriers (\citet{Her95},  \citet{Sno06}).

\begin{figure}
\epsscale{1}
\plotone{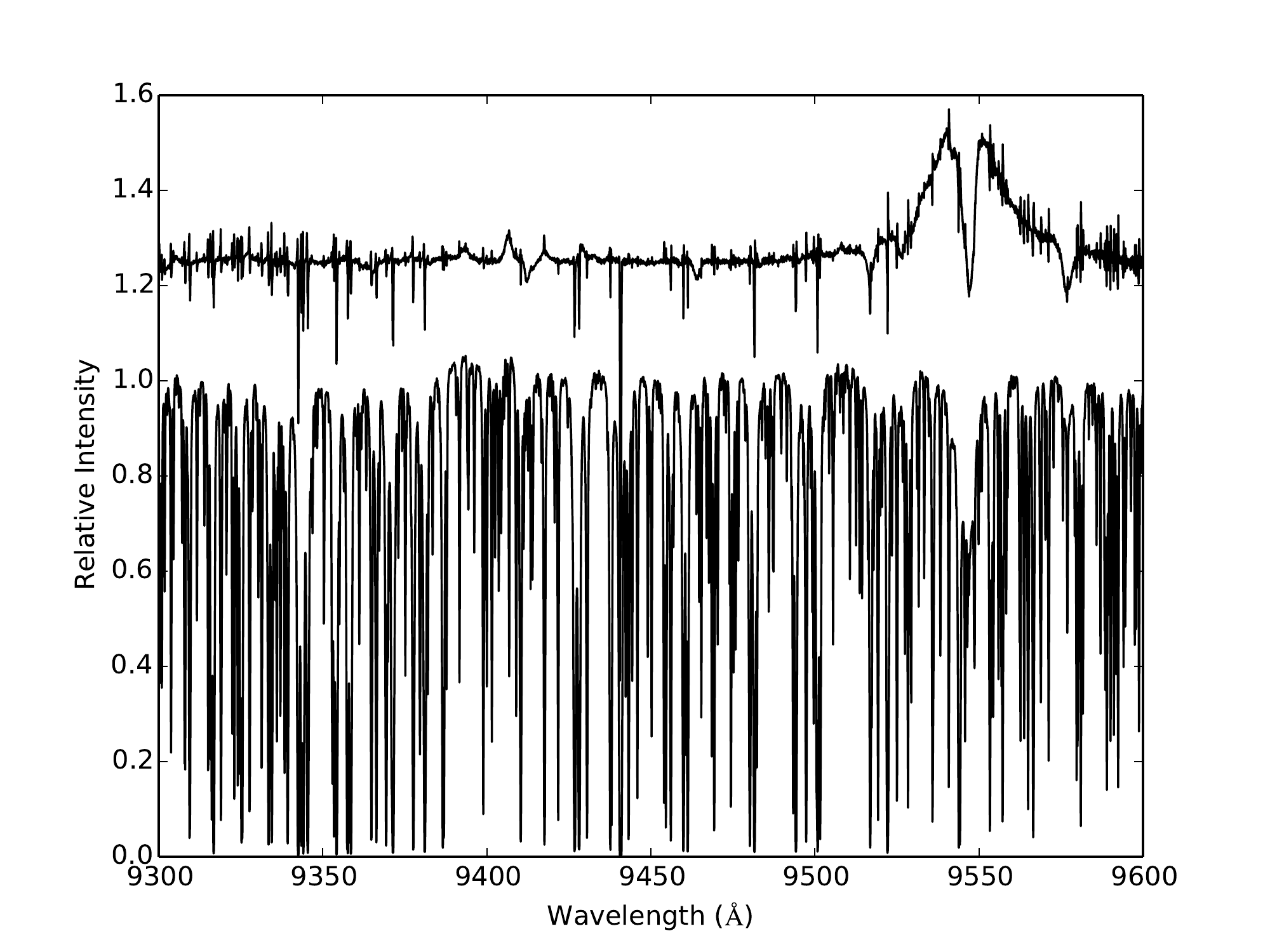}
\caption{ The lower spectrum of HD~183143 taken at CFHT contains the region of the C$_{60}^{+}$ absorption bands observed in the laboratory. The spectrum is dominated by strong telluric water vapour absorption lines. The upper plot follows division of the lower HD~183143 spectrum by that of HR~7437 which eliminates most of the water vapour lines. The feature at 9546\,\AA\/ is caused by hydrogen (P8), the narrow absorption component is from HD 183143 while the broad reflex (apparent emission) is from the broader P8 absorption line in HR 7437. Weak emission features from HD 183143 can be seen throughout the upper plot as well as a number of DIBs. The  C$_{60}^{+}$ 9577.5\,\AA\/ band is clearly visible.  \label{figure1} }
\end{figure}

\citet{Campbell2015} also reported two weaker bands of roughly equal strength (3:2) at 9428.5 and 9365.9\,\AA\/ in their laboratory spectra (their Table 1). Unfortunately, for ground based observations, both of these wavelengths lie in spectral regions heavily contaminated by telluric water vapour (WV) lines as can be seen in Figure \ref{figure1}.  The standard technique for their removal is division of the reddened star spectrum by one from an unreddened  star of similar spectral type observed through the same air mass.

In this paper, we report observations of HD 183143 chopped with those of a similar standard lying within $2^{\circ}$ on the sky.  Atmospheric WV absorption is likely to change with both time and direction as well as with air mass which is the reason for frequent chopping. HD 183143 is a bright, reddened, B7 supergiant which exhibits all of the known DIBs with considerable strength  \citep{Her95} including those at 9632 and 9577\,\AA\/. It also passes through the zenith at CFHT where the atmospheric precipitable water level can be significantly lower than for sites at lower altitude.

\section{Observations of HD 183143}

We were granted two hours of CFHT Director's Discretionary Time with ESPaDonS  \citep{Don2003} on the night of 2015-07-28 UT.  HD~183143 and a nearby standard of similar spectral type, HR~7437, were observed within one hour of meridian crossing with a sky temperature of $\sim\,-38\,^{\circ}\textrm{C}$ indicating a low water vapour level. ESPaDonS was configured in spectropolarimetric mode consequently, a single spectrum consisted of four individual sub-exposures taken at different polarimeter retarder configurations.   All of the spectra were reduced at CFHT with the Upena pipeline using the Libre-ESpRIT code \citep{Donati1997}.  The spectral resolution is approximately 65,000 or 0.05\,\AA\/ per pixel. Details of the stars are given in Table 1 which includes V and I magnitudes, and colour excess, $E_{(B-V)}$.

Observations alternated between the two targets.  In this way, four sets of spectra were acquired with air masses differing by $<\,0.01$ within each pair.  The total exposure times of 600 and $720\,\textrm{s}$ within each set for HD~183143 and HR~7437, respectively, were chosen to give similar signal levels for both stars in the 9400\,\AA\/ region. 

Telluric lines in typically 20\,\AA\/ regions of interest in the spectrum of HD~183143 were removed using the subsequent HR~7437 spectrum as standard in the IRAF\footnote{IRAF is distributed by the National Optical Astronomy Observatory, which is operated by the Association of Universities for Research in Astronomy, Inc., under cooperative agreement with the National Science Foundation.} telluric program.  The four corrected spectra of HD~183143 from each of these spectral regions were then combined to produce the single high S/N spectra in Figure \ref{figure2}. Estimation of S/N is complicated by the presence of the telluric lines within which S/N is low. Based on our calculations, S/N $\sim$ 1500 \AA$^{-1}$ ($\sim$ 330 per pixel) for HD~183143 in the region of the 9577.5\,\AA\/ line after WV removal.

There is a short gap in the ESPaDOnS spectral coverage between 9608 and 9636\,\AA\/ where the order lies off the detector.  This gap prevented our looking for the 9632.7\,\AA\/  band.

\begin{deluxetable}{clccc}
\tabletypesize{\footnotesize}
\tablecaption{CFHT ESPaDonS Observations \label{observations}}
\tablewidth{0pt}
\tablehead{
\colhead{star} & \colhead{Sp/L} & \colhead{I} & \colhead{V} & \colhead{$E_{(B-V)}$} \\}
\startdata
HD183143 & B7 Ia   & 4.79 & 6.86 & 1.28    \\
HR 7437& B8 IIIn  &5.10 & 5.00 & 0.00   \\ 
HD 169454 & B1 Ia & 5.13 & 6.71 & 1.12  \\
HD 177724 & A4 IV-Vnn & 2.98 & 2.99 & 0.00 \\
 \enddata
\end{deluxetable}

\section{Results}

The wavelength region of interest from 9300 to 9600\,\AA\/ is covered in a single \'echelle spectral order (24) and shown in the lower plot of Figure \ref{figure1} for HD 183143. All of the strong lines, some near saturation, are due to telluric WV.  The upper plot is the result of division by the spectrum of HR 7437. Note that the telluric line cancellation has not been optimized in the figure since in this case a single correction was applied to the full 300\,\AA\/ region displayed.  The narrow stellar hydrogen P8 line at 9546\,\AA\/ in HD~183143 is superimposed on the reflex of the much broader P8 line in HR~7437 illustrating the impact of absorption lines in the standard star. The strong 9577.5\,\AA\/  band of $\textrm{C}_{60}^{+}$ is easily seen. Several other DIBs are also visible. 

To optimize telluric line removal we performed these corrections over more limited spectral ranges containing the wavelengths of the $\textrm{C}_{60}^{+}$ laboratory absorptions.  
This produced higher S/N spectra for the 9577.5, 9428.5 and 9365.8\,\AA\/ regions as well as a region centered on the interstellar potassium line at 7699.0\,\AA\/ in order 29.

\begin{figure}
\epsscale{1}
\plotone{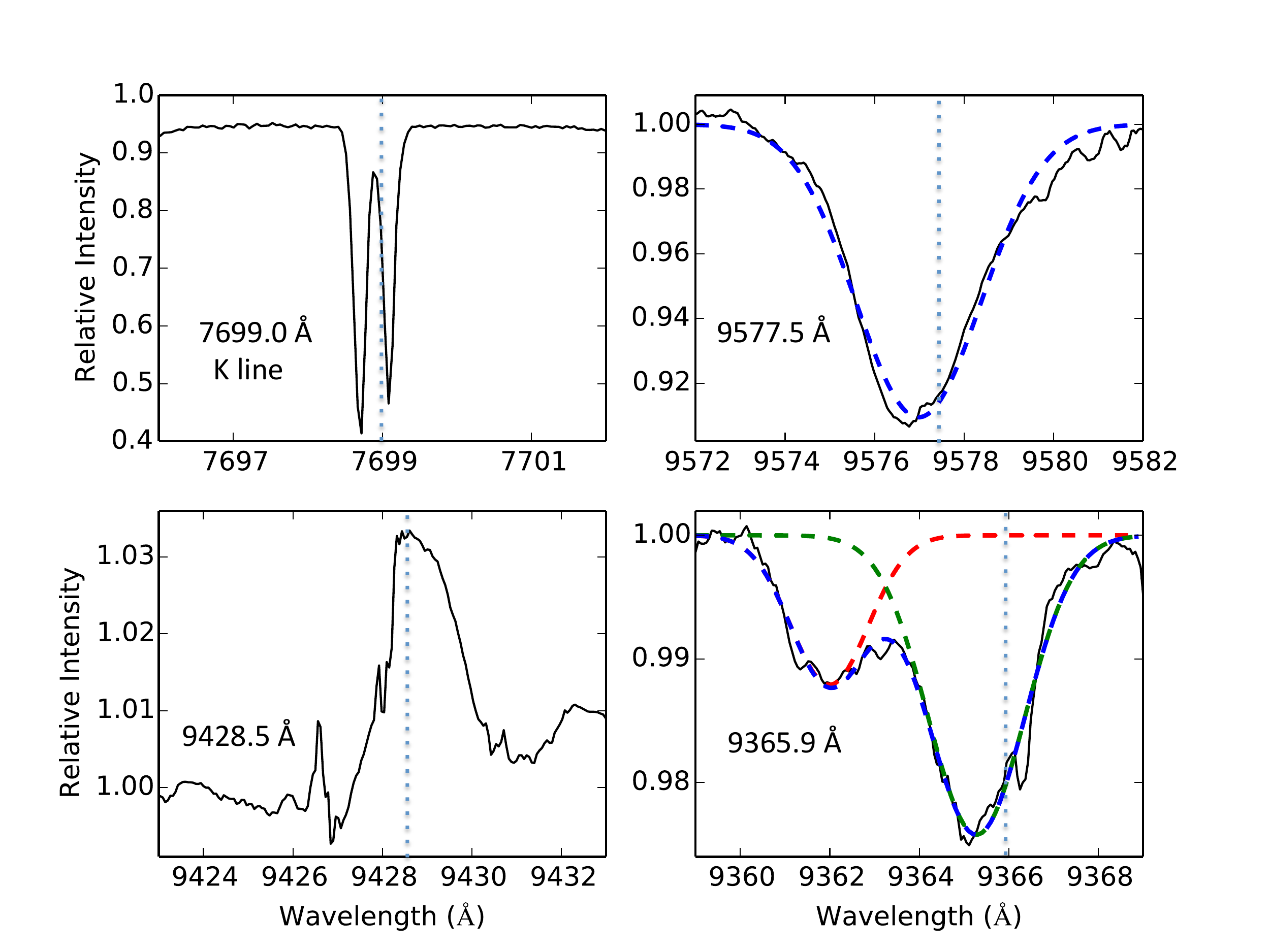}
\caption{The spectral regions of interest for HD~183143 after cancelation of telluric water vapour lines by division with HR~7437 and adjustment of zero levels. The vertical dotted lines indicate laboratory rest wavelengths for the 7699.0\,\AA\/ potassium line, and the absorption bands of $\textrm{C}_{60}^{+}$. The dashed lines are Gaussian fits for which parameters are listed in Table 2. The emission feature at 9429\,\AA\/ prevented any interstellar band detection for HD~183143. A  DIB  is also seen at 9362.5\,\AA\/ and fitted separately with a Gaussian. 
\label{figure2} }
\end{figure}

Small zero point adjustments were made to the levels of the HD~183143 and HR~7437 spectra to minimize residuals in the ratios at the centres of the strongest WV lines. The ratios are shown for the four regions in Figure \ref{figure2}. The persistence of some residuals at the centres of the strongest telluric lines is likely associated with non-linearity of the analog to digital converter over the extreme range of intensity of the lines rather than inequality of water vapour optical depth.

Figure \ref{figure2} also shows the two components of the HD~183143 interstellar 7699.0\,\AA\/ potassium absorption line after removal of nearby telluric lines. The stronger component has a radial velocity of $-$12 km s$^{-1}$, the other component  +4 km s$^{-1}$.  At 9577.5, 9428.5 and 9365.8\,\AA\/, $-12\,\textrm{km}\,\textrm{s}^{-1}$ corresponds to a displacement of $-$0.4\,\AA\/. We use this value in what follows assuming that the corresponding  cloud is the main contributor to the detected DIBs. 

\citet{Campbell2015} fitted Gaussians to the laboratory absorptions at $6\,\textrm{K}$ and listed wavelengths and FWHMs (their Table 1). The DIBs in Figure \ref{figure2} were also fitted with unconstrained Gaussians  as shown by the dashed lines including the obvious DIB at $9362.5\,\textrm{\AA}$. No DIBs could be observed in the region of the $9428.5\,\textrm{\AA}$ laboratory absorption because of the emission feature at 9429\,\AA\/. The emission must be associated with HD~183143 because the comparison star, HR~7437, is a rapid rotator ($v \sin i\,=\,265\,\textrm{km}\,\textrm{s}^{-1}$) such that absorption lines in HR~7437 would produce much  broader ($>$ 8\AA\/) apparent emission features in the ratio. There are many such emission lines between 9300 and 9600\,\AA\/ for HD~183143. \citet{Jenniskens1997}  published spectra in this region, and encountered emission lines in either their reddened or standard stars (their Figure 4). It is interesting to note that in the latter study, which was published long before the gas phase spectrum of C$_{60}^{+}$ was recorded, both a `depression' near $9428\,\textrm{\AA}$ and the suggestion of a weak feature around $9366\,\textrm{\AA}$ were reported.

\begin{figure}
\epsscale{1}
\plotone{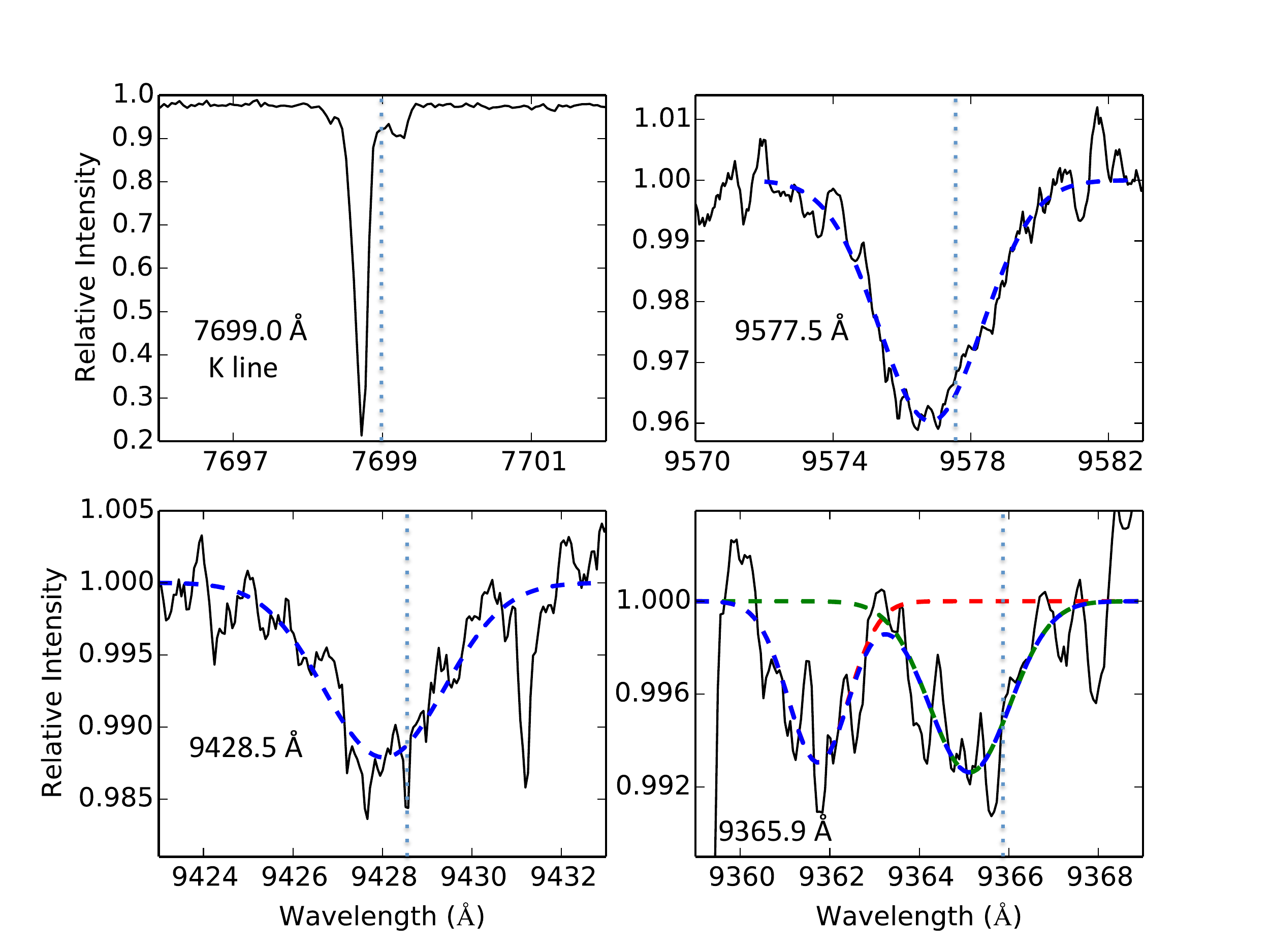}
\caption{ The same spectral regions  for HD~169454 after cancelation of telluric water vapour lines by division with HD~177724 and adjustment of zero levels. The vertical dotted lines indicate laboratory rest wavelengths for the 7699.0\,\AA\/ potassium line, and the absorption bands of $\textrm{C}_{60}^{+}$. The dashed lines are Gaussian fits for which parameters are listed in Table~2.  In the panel for 9428.5\,\AA\/, large water vapour line residuals  at 9427.5 and 9428.2\,\AA\/ have been omitted and replaced by averages. The DIB  seen at 9362.5\,\AA\/ is fitted with a separate Gaussian.  \label{figure3}}
\end{figure}

Figure \ref{figure3} shows the regions of the 9577.5, 9428.5 and 9365.9\,\AA\/ laboratory bands for the B1 Ia star HD~169454 divided by the spectrum of the rapidly rotating ($v \sin i\,=\,317\,\textrm{km}\,\textrm{s}^{-1}$) AO V star HD~177724, again using the IRAF telluric program.  Details of the stars are given in Table 1. These processed spectra were downloaded from the Canadian Astronomy Data Centre's CFHT archive. Both stars were observed in the ``object  only'' spectroscopic configuration of ESPaDonS on 2005-05-21 UT.  We selected HD 169454 because it has none of the emission features found in HD 183143. Unfortunately, the standard was not observed at an identical air mass and so there are  large WV line residuals at  9427.5 and 9428.2\,\AA\/ which have been omitted and replaced by averages in the plots and when fitting a Gaussian. The principal interstellar K line (7699.0\,\AA\/) is unresolved with a radial velocity of $-$11 km~s$^{-1}$ which translates to $-$0.4\,\AA\/ at 9400\,\AA\/. Table~2 summarises the results of unconstrained Gaussian fits to the features in all of the ratioed spectra. 

Given the weakness ($\sim1\%$) of the detected bands and the presence of apparent emission features, setting of  continua was subjective, particularly in the case of HD~169454. Further, the CCD detector is partially transparent at such long wavelengths which generates interference fringes. These are not well cancelled in the flat fielding process but the additional normalization when dividing by the telluric standard considerably reduces their amplitude. Consequently, it is hard to establish values for systematic errors with confidence particularly for the FWHM in the case of HD~169454. In the table, only formal fitting errors are given.

\section{Comparison with C$_{60}^{+}$  laboratory bands}

The agreement of the astronomical and laboratory wavelengths is remarkably good considering the somewhat ad hoc nature of assigning an interstellar velocity. In addition, support for the assignment to $\textrm{C}_{60}^{+}$ comes from comparison with the relative intensities reported in the laboratory measurements. The depth of the astronomical bands (Table 2) can be used to estimate the relative intensity of the absorption bands for symmetric profiles. The observations towards HD~183143 give a ratio of 1:0.3 for the 9577.5 and 9365.9\,\AA\/ bands. This matches with the laboratory ratio of 1:0.2 which was reported with an estimated uncertainty of around 20\,\% \citep{Campbell2015}. Furthermore, the ratio of the depth values derived from the archive HD~169454 spectra, 1:0.3:0.2 for the 9577.5, 9428.5 and 9365.9\,\AA\/ DIBs, are also in agreement with 1:0.3:0.2 measured in the laboratory \citep{Campbell2015}.

The FWHM of 3.3\,\AA\/ for the interstellar 9577.5\,\AA\/ band in HD~183143 is comparable to the value of 2.85\,\AA\/ reported by \citet{Foi97} and 3.0\,\AA\/ by \citet{Jenniskens1997} for other reddened stars. In the laboratory study at $6\,\textrm{K}$ the measured FWHM are $2.5\,\textrm{\AA}$. At this low temperature, the FWHM of the $\textrm{C}_{60}^{+}$ rotational profile is around 1\,\AA\/ \citep{Edw93}. This indicates that the laboratory bands are broadened by the $2\,\textrm{ps}$ lifetime of the excited electronic state. Only at temperatures above 30 K would one see a further broadening of the rotational profile. The FWHM values for HD~169454 are not robust because of incomplete correction for telluric contamination and arbitrariness in setting the continuum. A more likely source for the extra broadening of the 9577.5\,\AA\/ band in HD~183143 would be near coincidence with another DIB. For example, there is a close DIB to the 9365.9\,\AA\/ band at 9362.5\,\AA\/ (Figure \ref{figure2}).

\section{Astromomical implications}

The confirmation of the presence of $\textrm{C}_{60}^{+}$ in diffuse clouds raises the question of its interstellar abundance. This has been considered previously, see, for example \citep{Her00}, and references therein. The calculation of column density relies on the knowledge of the individual band oscillator strength, $f(\lambda)$, the wavelength of the absorption band maximum, $\lambda_c$, and the equivalent width, $W$. For this purpose the value for the electronic transition which was estimated in the matrix isolation spectroscopy study by \citet{Ful93}, $f_{e}\,=\,0.003-0.006$, has invariably been used. However, this value is an order of magnitude smaller than indicated by theory \citep{Ben12}. We believe that the value from the matrix study is underestimated and in the following $f_{e}\,=\,0.05$ is used. 

In order to evaluate the $f(\lambda)$ value for the $9577.5\,\textrm{\AA\/}$ band the electronic oscillator strength must be weighted by the Franck-Condon factor. In the case of $\textrm{C}_{60}^{+}$, most of the oscillator strength is localized in the two strong features at $9632.7\,\textrm{\AA}$ and $9577.5\,\textrm{\AA}$ \citep{Campbell2015}. Based on observation of only weak $\textrm{C}_{60}^{+}$ laboratory absorptions to shorter wavelength than $9428.5\,\textrm{\AA}$, we estimate the 9577.5\,\AA\/ band to account for $\sim\,30\,\%$ of the intensity and therefore obtain $f(9577.5\,\textrm{\AA\/})\,\simeq\,0.02$. For the 9577.5\,\AA\/ DIB observed towards HD~183143 the equivalent width is estimated to be $W\,\sim\/\,0.3\,\textrm{\AA\/}$ (Table 2). Using these values gives a column density, $N(\textrm{C}_{60}^{+})$, of around $2\,\times\,10^{13}\,\textrm{cm}^{-2}$. For comparison, this is an order of magnitude smaller than that of $\textrm{H}_{3}^{+}$ but similar to $\textrm{CH}^{+}$, both of which were observed towards HD~183143 \citep{McC02}.

\section{Conclusion and outlook}

Two weak diffuse interstellar bands have been detected with wavelengths, FWHM and relative intensities in agreement with the laboratory spectrum of $\textrm{C}_{60}^{+}$ \citep{Campbell2015}. These results provide further compelling evidence for the presence of this molecular ion in the interstellar medium. It has been suggested that the $9632.7\,\textrm{\AA\/}$ and $9577.5\,\textrm{\AA\/}$ absorptions, and hence $\textrm{C}_{60}^{+}$, are also present in protoplanetary nebulae \citep{Gro13}.  The existence of the neutral fullerenes $\textrm{C}_{60}$ and $\textrm{C}_{70}$ in a young planetary nebula \citep{Cam2010} and reflection nebulae \citep{Sell2010} has been confirmed through their infrared (IR) transitions. There is also a report of IR emission features from a reflection nebula which have been proposed to be due to $\textrm{C}_{60}^{+}$  \citep{Bern2013}. The identification of $\textrm{C}_{60}^{+}$ in diffuse clouds leads to intriguing questions regarding the role of fullerenes with respect to the formation of smaller carbon based molecules such as those identified in dense interstellar clouds by radioastronomy.

\begin{deluxetable}{ccc|ccc|ccc}
\tabletypesize{\footnotesize}
\tablecaption{ Laboratory and Interstellar C$_{60}^{+}$ Bands\tablenotemark{\#}  \label{comparison} }
\tablewidth{0pt}
\tablehead{
\multicolumn{3}{c}{laboratory} &
\multicolumn{3}{c}{HD~183143} &
\multicolumn{3}{c}{HD~169454} \\
\colhead{$\lambda_c$  \AA}& 
\colhead{$\sigma_{\rm rel}$\tablenotemark{\dag}} & \colhead{FWHM \AA} &
\colhead{$\lambda_c$\tablenotemark{\ddag}  \AA }&
\colhead{depth \%} & \colhead{FWHM \AA } &
\colhead{$\lambda_c$\tablenotemark{\ddag}  \AA  }&
\colhead{depth \%} & \colhead{FWHM  \AA} 
}
\startdata
9577.5 $\pm$0.1& 1   & 2.5 $\pm$0.2 & 9577.4 $\pm$0.02& 9.1 $\pm$0.05 & 3.3 $\pm$0.04&9577.2 $\pm$0.03& 4.0 $\pm$0.06   & 3.5 $\pm$0.06   \\
9428.5 $\pm$0.1& 0.3  & 2.4 $\pm$0.2& - & - &- & 9428.4 $\pm$0.1&1.2 $\pm$0.05 & 3.2 $\pm$0.1 \\ 
9365.9 $\pm$0.1& 0.2    & 2.4 $\pm$0.2 & 9365.7 $\pm$0.02 &2.4 $\pm$0.03 & 2.5 $\pm$0.04 &9365.6 $\pm$0.1& 0.7 $\pm$0.1  & 2.1 $\pm0.2$\\
 \enddata
\tablenotetext{\#}{central wavelengths ($\lambda_c$), depths, FWHM and errors from Gaussian fits to the lab and astronomical bands.  No systematic errors included.}
\tablenotetext{\dag}{relative cross section}
\tablenotetext{\ddag}{corrected by +0.4\,\AA\/  for the interstellar K line off-set}

\end{deluxetable}

\acknowledgements
The authors thank the Director of CFHT, Dr Simons, for the prompt assignment of telescope time and the highly professional quality of the observations made by the CFHT staff. This research used the facilities of the Canadian Astronomy Data Centre operated by the National Research Council of Canada with the support of the Canadian Space Agency.

\begin{thebibliography}{}

\bibitem[Bendale et al.(1992)]{Ben12} Bendale, ~R. ~D., Stanton, ~J. ~F. and Zerner, ~M. ~C. 1992, Chem. Phys. Lett., 194, 467

\bibitem[B\'erne et al.(2013)] {Bern2013} B\'erne,~O., Mulas,~G. and Joblin,~C. 2013, \aap, 550, L4

\bibitem[Campbell et al.(2015)]{Campbell2015} Campbell, ~E.~K., Holz,~M., Gerlich,~D. and Maier,~J.~ P. 2015,
\nat, 523, 322 

\bibitem[Cami et al.(2010)]{Cam2010}  Cami,~J., Bernard-Salas,~J., Peeters,~E. and Malek,~S.~E. 2010, Science, 329, 1180

\bibitem[Donati(2003)]{Don2003}Donati,~J.-F. 2003, in ASP Conf. Ser. 307, Solar Polarization, ed. J. Trujillo-Bueno and J. Sanchez Almeida (San Francisco: ASP), 41

\bibitem[Donati et al.(1997)]{Donati1997} Donati,~J.-F., Semel,~M., Carter,~B.~D., Rees,~D.~E., Cameron,~A.~C., 1997, \mnras, 291, 658

\bibitem[Edwards and Leach(1993)]{Edw93} Edwards,~S.~A. and Leach,~S. 1993,  \aap, 272, 533

\bibitem[Foing and Ehrenfreund(1994)]{Foing1994} Foing,~B.~H. and Ehrenfreund,~P. 1994, \nat, 369, 296

\bibitem[Foing and Ehrenfreund(1997)]{Foi97} Foing,~B.~H. and Ehrenfreund,~P. 1997, \aap, 317, L59

\bibitem[Fulara et al.(1993)]{Ful93} Fulara,~J.,  Jakobi,~M. and Maier,~J.~P. 1993, Chem. Phys. Lett., 211, 227

\bibitem[Herbig(1995)]{Her95} Herbig,~G.~H., 1995, \araa, 33, 19

\bibitem[Herbig(2000)]{Her00} Herbig,~G.,~H., 2000,  \apj, 542, 334

\bibitem[Iglesias-Groth and Esposito(2013)]{Gro13} Iglesias-Groth,~S. and Esposito,~M. 2013, \apj, 776, L2

\bibitem[Jenniskens et al.(1997)]{Jenniskens1997}Jenniskens,~P., Mulas,~G., Porceddu,~I. and Benvenuti, P. 1997, \aap, 327, 337

\bibitem[Kroto(1987)]{Kro87} Kroto,~H.~W., 1987, Polycyclic Aromatic Hydrocarbons and Astrophysics, eds. A. L\'eger et al. (Riedel Publishing Company), 197

\bibitem[McCall et al.(2002)]{McC02} McCall,~B.~J. Hinkle,~K.~H., Geballe,~G.~H., Moriarty-Schieven,~G.~H., Evans II,~N.~J., Kawaguchi,~K., Takano,~S. Smith,~V.~V. and Oka,~T. 2002, \apj, 567, 391

\bibitem[Sellgren et al.(2010)]{Sell2010} Sellgren,~K., Werner,~M.~W., Ingalls,~J.~G., Smith,~J.~G.~T., Carleton,~T.~M., and Joblin,~C. 2010 \apj, 722, L54

\bibitem[Snow and McCall(2006)]{Sno06}Snow,~T.~P. and McCall,~B.~J.  2006, \araa, 44, 367

\end{thebibliography}
\end{document}